\documentclass[aps,prl,twocolumn,amsfonts,amsmath,amssymb,showpacs,final,10pt]{revtex4-1}
\usepackage[dvipsnames]{xcolor}
\usepackage[colorlinks=true,citecolor=Blue,linkcolor=Blue,urlcolor=Blue]{hyperref}
\usepackage{graphicx,manfnt,pifont,colonequals}
\usepackage{bbm}

\newcommand\noi{\noindent}
\newcommand\Tr{{\rm Tr}}

\newcommand\bM{{\mathbb M}}

\newcommand\cB{{\mathcal B}}
\newcommand\cC{{\mathcal C}}

\newcommand\cM{{\mathcal M}}
\newcommand\cN{{\mathcal N}}
\newcommand\cO{{\mathcal O}}
\newcommand\cR{{\mathcal R}}
\newcommand\cS{{\mathcal S}}
\newcommand\cT{{\mathcal T}}
\newcommand\cV{{\mathcal V}}

\newcommand\ef{\mathfrak{e}}
\newcommand\ff{\mathfrak{f}}
\newcommand\gf{\mathfrak{g}}
\newcommand\suf{\mathfrak{su}}
\newcommand\uf{\mathfrak{u}}
\newcommand\sof{\mathfrak{so}}

\newcommand\uspf{\mathfrak{usp}}

\newcommand\restr[2]{{\left.\kern-\nulldelimiterspace#1\vphantom{\big|}\right|_{#2}}}
\newcommand\ie{\emph{i.e.}}

\begin{document}
\title{Flavor symmetries and unitarity bounds in \texorpdfstring{$\cN=2$}{N=2} SCFTs}
\author{Christopher Beem} 
\affiliation{Mathematical Institute, University of Oxford, United Kingdom}

\begin{abstract}
\noi In this letter I analyze the constraints imposed by unitarity on the flavor central charges of four-dimensional $\cN=2$ SCFTs with general reductive global symmetry groups. I derive several general and far-reaching consequences of unitarity by computing the norms of flavor singlet Higgs branch operators appearing in the squares of ``moment map'' operators via the associated vertex operator algebra, and imposing the requirement that they be non-negative.
\end{abstract}

\maketitle

\section{Introduction}
\label{sec:intro}

The detailed structure of any four-dimensional $\cN=2$ SCFT is constrained by the existence of a vertex operator algebra (VOA) structure \cite{Beem:2013sza} on the space of Schur operators \cite{Gadde:2011uv}. In this letter, I will investigate the interplay between four-dimensional unitarity and the VOA structures associated to flavor symmetries. This subject was previously considered to some extent in the original work of \cite{Beem:2013sza}, as well as in the subsequent work \cite{Lemos:2015orc}, while similar issues were investigated in \cite{Beem:2013qxa,Beem:2016wfs,Liendo:2015ofa}. However, as we shall see, there is more left to say.

The basic fact that we will exploit is that if a four-dimensional SCFT $\cT$ has a continuous flavor symmetry $\gf$, then its associated VOA $\chi[\cT]$ includes as a vertex operator subalgebra a $\hat{\gf}_{k}$ affine Kac-Moody VOA,
%%%%%%
\begin{equation}\label{eq:KM_VOA_OPE}
J^a(z)J^b(w)\sim \frac{kd^{ab}}{(z-w)^2}+\frac{i f^{ab}_{\phantom{ab}c}\,J^c(w)}{z-w}~.
\end{equation}
%%%%%%
where the Kac-Moody level $k$ is related to the flavor central charge $k_{4d}$ in $\cT$ according to $k = -\frac12 k_{4d}$, and where $d_{ab}=\Tr(T_aT_b)$ is the Killing form on $\gf$, normalized so that long roots have squared length equal to two. Similarly, locality of the theory $\cT$ (expressed through the existence of a stress tensor multiplet whose charge algebra generates superconformal transformations) implies that $\chi[\cT]$ includes as a vertex operator subalgebra a ${\rm Vir}_{c}$ Virasoro VOA, 
%%%%%%
\begin{equation}\label{eq:Vir_VOA_OPE}
T(z)T(w)\sim \frac{\frac{c}{2}}{(z-w)^4}+\frac{2T(w)}{(z-w)^2}+\frac{T'(w)}{z-w}~.
\end{equation}
%%%%%%
where the Virasoro central charge is related to the Weyl anomaly coefficient $c_{4d}$ of $\cT$ according to $c = -12c_{4d}$. 

What will be important to us is that the detailed structures of these subalgebras are completely determined in terms of the coefficients $c$ and $k$, and these detailed structures include the two- and three-point functions of (infinitely many) operators, each of which corresponds to some definite local operator in $\cT$. To the extent that the identification between two- and four-dimensional operators can be made precise, one therefore has in principle an infinite number of non-trivial constraints that must be obeyed in order for $\cT$ to be a unitary SCFT. It is, however, in general quite a non-trivial task to disambiguate the identification of specific four-dimensional operators in the VOA. 

It turns out that in the presence of continuous flavor symmetries, the map between VOA states and four-dimensional operators can be understood exactly for a specific class of states, namely, states of the form $d_{ab}J^a_{-1}J^b_{-1}\Omega$. This will allow us to determine the exact matrix of inner products of a key class of Higgs branch chiral ring operators, from which will follow a number of nontrivial unitarity bounds relating to flavor symmetries and central charges.

%%%%%%%%%%%%%%%%%%%%%%%%%%%%%%%%%%%%%%%%%%%%%%%%%%%%%%%%%%%%%%%%%%%%%%%%%%%%%%%%%%%%%%%%%%%%%%%%%%%%%%%%%%%%%%%%%%%%%
%%%%%%%%%%%%%%%%%%%%%%%%%%%%%%%%%%%%%%%%%%%%%%%%%%%%%%%%%%%%%%%%%%%%%%%%%%%%%%%%%%%%%%%%%%%%%%%%%%%%%%%%%%%%%%%%%%%%%

\vspace{-4pt}
\section{Review of simple flavor bounds} 
\label{sec:single_factor_review}

In \cite{Beem:2013sza}, a simple inequality was derived between the $c$ central charge and the level $k$ associated to a simple factor $\gf$ of the global flavor symmetry group in any unitary theory with a unique stress tensor multiplet. In particular, the constraints of unitarity for the four-point function of conserved current multiplets were shown to imply that these central charges must satisfy
%%%%%%
\begin{equation}\label{eq:single_factor_bound}
kd_{\gf} \leqslant c(k+h^\vee)~,
\end{equation}
%%%%%%
where $d_{\gf}$ and $h^\vee$ are the dimension and dual Coxeter number of $\gf$, respectively. When this bound is saturated, one has
%%%%%%
\begin{equation}\label{eq:single_channel_sugawara_saturation}
c = c_{\rm sug} = \frac{kd_{\gf}}{k+h^\vee} ~,
\end{equation}
%%%%%%
where $c_{\rm sug}$ represents the central charge associated to the Sugawara stress tensor of the affine Kac-Moody VOA,
%%%%%%
\begin{equation}\label{eq:single_channel_sugawara_stress_tensor}
T_{\rm sug} = \frac{d_{ab}(J^aJ^b)}{2(k+h^\vee)}~.
\end{equation}
%%%%%%
A further consequence of the four-point function analysis is that when the bound \eqref{eq:single_factor_bound} is saturated, the (VOA image of the) unique stress tensor multiplet of the four-dimensional theory, which we shall denote by $T_{\cT}$, must be identical to the Sugawara stress tensor in the affine Kac-Moody VOA,
%%%%%%
\begin{equation}\label{eq:sug_is_sug}
T_{\cT} \equiv T_{\rm sug}~.
\end{equation}
%%%%%%

It will be useful for us to revisit \eqref{eq:single_factor_bound} in a slightly different light. Given a (simple or abelian) affine Kac-Moody sub-VOA, one can form the (un-normalized) Segal-Sugawara operator 
%%%%%%
\begin{equation}\label{eq:SS_definition}
S = d_{ab}J^a_{-1}J^b_{-1}\Omega~.
\end{equation}
%%%%%%
The currents $J^a$ correspond to ``moment map operators'' $\mu^a$ in the Higgs branch chiral ring, which lie in $\widehat{\cB}_1$ multiplets in four dimensions \footnote{We employ the naming conventions of \cite{Dolan:2002zh} for unitary multiplets of the four-dimensional $\cN=2$ superconformal algebra.}. Let us also introduce the VOA operator $\mu^2$ corresponding to the flavor singlet appearing in the (Higgs chiral ring) square of the moment map,
%%%%%%
\begin{equation}\label{eq:moment_maps_VOA_ops}
J^a = \chi(\mu^a)~,\quad \mu^2 = \chi(d_{ab}\mu^a\mu^b)~,
\end{equation}
%%%%%%
where we represent the VOA image of a Schur operator $\cO$ by $\chi(\cO)$. 

Crucially, the normally-ordered product in the VOA is a different operation than the Higgs branch chiral multiplication, so in general $\mu^2\neq S$. However, four-dimensional selection rules imply that the operator $S$ must correspond to a linear combination of $\mu^2$, which lies in a $\widehat{\cB}_2$ multiplet, and the $\suf(2)_R$ current in the \emph{unique} $\widehat{\cC}_{0(0,0)}$ multiplet (\ie, stress tensor multiplet). In other words, we can re-write the Segal-Sugawara operator as
%%%%%%
\begin{equation}\label{eq:SS_decomp}
S = \mu^2 + \alpha\,T_{\cT}~,
\end{equation}
%%%%%%
where now each term on the right hand side represents the VOA image of a four-dimensional operator in a distinct representation of the $\suf(2,2|2)$ superconformal algebra.

Importantly, because they arise from four-dimensional operators living in different superconformal multiplets, the VOA inner product $\langle\mu^2|T_{\cT}\rangle=0$. Thus the constant $\alpha$ can be determined by considering the inner product $\langle T_\cT|S\rangle = \langle T_\cT | \alpha T_\cT\rangle$, where the left hand side is fixed by the requirement that the $J^a$ are Virasoro primary operators. With our present conventions, this gives
%%%%%%
\begin{equation}\label{eq:T_coefficient_fixed}
\alpha = \frac{2kd_{\gf}}{c}~.
\end{equation}
%%%%%%
By the same reasoning, we can further compute the VOA norm of the Higgs branch operator $\mu^2$ unambiguously,
%%%%%%
\begin{equation}\label{eq:quadratic_higgs_norm}
\begin{split}
|\!|\mu^2|\!|^2 &= |\!|S|\!|^2 - \alpha^2|\!|T_{\cT}|\!|^2~,\\
&=2 (k+h^\vee)kd_{\gf}-\tfrac{1}{2}\alpha^2c~,\\
&=2 kd_{\gf}\left((k+h^\vee)-\tfrac{k}{c}d_{\gf}\right)~.\\
\end{split}
\end{equation}
%%%%%%

In general, four-dimensional unitarity does not require VOA norms to be positive. For (four-dimensional) scalar Schur operators, the twisted-translation prescription of \cite{Beem:2013sza} implies that one must consider the order-four automorphism $\sigma:\cV\to\cV$ on the vector space $\cV$ underlying $\chi[\cT]$ that takes an operator to the complex conjugate of the $\suf(2)_R$ lowest weight state in the same multiplet \footnote{This conjugation operation is analogous to the conjugation on Higgs branch operators used in \cite{Gaiotto:2008nz} to describe the hyperk{\"a}hler structure of the Higgs branch, but supplemented with an extra complex conjugation and generalized to include operators with half-integer $R$-charge.}, \ie,
%%%%%%
\begin{equation}\label{eq:conjugation_operation}
\sigma\circ\cO = (-1)^{2R}\left(e^{\pi i\cR_{2}}\cO e^{-\pi i\cR_{2}}\right)^\ast~.
\end{equation}
%%%%%% 
The four-dimensional norms are then given in terms of VOA inner products by
%%%%%%
\begin{equation}\label{eq:4d_norms}
\begin{split}
|\!|\psi|\!|^2_{4d} &= \langle\cO|\sigma\circ\cO\rangle~.\\
				    &= z^{2h_{\cO}}\langle\cO(z)(\sigma\circ\cO)(0)\rangle~.
\end{split}
\end{equation}
%%%%%%
(For operators with spin, the procedure is marginally more complicated, but will be discussed, for example, in future work \cite{unitarity}). 

For the moment map operator, $R=1$ and conjugation acts as $\sigma\circ\mu^a = -\mu^a$ (here we take $T_a$ to be a real basis for $\gf$), and for the $\mu^2$ operator with $R=2$ it acts as $\sigma\circ\mu^2 = \mu^2$. Thus for this state, the VOA norm is required by four-dimensional unitarity to be non-negative. Because $k$ and $c$ are negative definite, this is equivalent to the bound \eqref{eq:single_factor_bound}. When the bound is saturated, the Higgs branch operator $\mu^2$ is identically zero, and \eqref{eq:SS_decomp} yields the identification \eqref{eq:sug_is_sug} of the Sugawara stress tensor with the true stress tensor.

%%%%%%%%%%%%%%%%%%%%%%%%%%%%%%%%%%%%%%%%%%%%%%%%%%%%%%%%%%%%%%%%%%%%%%%%%%%%%%%%%%%%%%%%%%%%%%%%%%%%%%%%%%%%%%%%%%%%%
%%%%%%%%%%%%%%%%%%%%%%%%%%%%%%%%%%%%%%%%%%%%%%%%%%%%%%%%%%%%%%%%%%%%%%%%%%%%%%%%%%%%%%%%%%%%%%%%%%%%%%%%%%%%%%%%%%%%%

\vspace{-4pt}
\section{Reductive flavor bounds}
\label{sec:reductive}

The analysis of \cite{Beem:2013sza}, recounted and reinterpreted above, was restricted to the case where a single simple (or ${\uf}(1)$) factor of the global flavor symmetry group was considered. However, given the explicit construction of flavor-singlet $\widehat{\cB}_2$ operators we have found, it is simple to generalize the above discussion to the case of a general reductive flavor group,
%%%%%%
\begin{equation*}\label{eq:reductive_symmetry}
\gf_F = \gf_1\times\cdots\times \gf_{n}~,
\end{equation*}
%%%%%%
with each factor $\gf_i$ either simple or $\uf(1)$. To assist in the discussion of several flavor factors at once, we adopt the convention of referring to a simple or $\uf(1)$ factor of the flavor group as
%%%%%%
\begin{itemize}
\item \emph{sub-critical} if $k + h^\vee > 0$ \quad(\ie, $k_{4d} < 2h^\vee$).
\item \emph{critical} if $k + h^\vee = 0$ \quad(\ie, $k_{4d} = 2h^\vee$).
\item \emph{super-critical} if $k + h^\vee < 0$ \quad(\ie, $k_{4d} > 2h^\vee$).
\end{itemize}
%%%%%%
With $k < 0$, these three cases correspond to $c_{{\rm sug}}$ being negative, zero, and positive, repectively, and by convention the dual Coxeter number of ${\uf}(1)$ is taken to be zero. Thus the bound \eqref{eq:single_factor_bound} has force only when considering simple flavor symmetry groups that are sub-critical.

In addition to the norms computed as above for each simple or $\uf(1)$ factor, there is no obstacle to analyzing the full matrix of inner products of $\mu^2_i$ states. Indeed, since the Segal-Sugawara vectors of distinct factors of the flavor group are orthogonal in the VOA, we have for $i\neq j$,
%%%%%%
\begin{equation}\label{eq:quadratic_higgs_overlap}
\langle \mu^2_i|\mu^2_j\rangle = -\frac{c\,\alpha_i\alpha_j}{2} = -\frac{2k_{i}k_{j}d_{{\gf}_i} d_{{\gf}_j}}{c}~.
\end{equation}
%%%%%%
This implies a simple expression for the complete matrix of inner products of quadratic Higgs branch singlets (\ie, the $\mu_i^2$ operators), which we denote by ${\mathbb M}$,
%%%%%%
\begin{equation}\label{eq:norm_matrix_structure}
{\bM} = \Delta - \frac{c}{2}\left(\vec{\alpha}\otimes\vec{\alpha}\right)~,
\end{equation}
%%%%%%
where we have
%%%%%%
\begin{equation}\label{eq:minor_ingredients}
\begin{split}
\Delta 			&= {\rm diag}\left(\delta_1,\cdots,\delta_n\right)~,\quad \delta_i = 2k_id_i(k_i+h_i)~,\\
\vec{\alpha} 	&= \left(\alpha_1,\ldots,\alpha_n\right)~,\qquad~~ \alpha_i = \tfrac{2k_id_i}{c}~.
\end{split}
\end{equation}
%%%%%%
It is precisely due to the mixing between the $\widehat{\cB}_2$ and $\widehat{\cC}_{0(0,0)}$ multiplets in the Segal-Sugawara vectors that this matrix has off-diagonal terms between $\mu^2_i$ states coming from different factors of the flavor group. For this reason, an analysis of the present type would not yield any new results beyond the simple case for quadratic states in non-singlet representations of the flavor group, since there can be no flavor-charged $\widehat{\cC}_{0(0,0)}$ multiplets under the present assumptions.

Four-dimensional unitarity requires that any linear combination of the $\mu^2_i$ have positive norm, which due to \eqref{eq:4d_norms} implies that the total matrix $\bM$ must be positive semi-definite (with zero eigenvalues corresponding to Higgs chiral ring relations amongst the quadratic Higgs branch singlets). To ensure positive semidefiniteness it is enough to ensure that all principal minors are non-negative, with the previous bound \eqref{eq:single_factor_bound} corresponding to $1\times 1$ principal minors, \ie, entries on the diagonal. However, the non-zero off-diagonal terms will give stronger unitarity constraints than those arising from the single-factor analysis. In what follows, we will derive the complete set of physical consequences of positive semi-definiteness of $\bM$. 

%%%%%%%%%%%%%%%%%%%%%%%%%%%%%%%%%%%%%%%%%%%%%%%%%%%%%%%%%%%%%%%%%%%%%%%%%%%%%%%%%%%%%%%%%%%%%%%%%%%%%%%%%%%%%%%%%%%%%

\vspace{-6pt}
\subsection{Non-critical factors}
\label{sec:sugawara_saturation}

To begin, let us restrict our attention to principal minors that don't include any rows/columns corresponding to critical flavor factors from $\gf_F$ (if any are present at all). For a collection of flavor factors $\gf_{i_1}\times\cdots\times\gf_{i_k}$, the corresponding minor takes the simple form
%%%%%%
\begin{equation}\label{eq:determinant_bound}
\begin{split}
{\rm det}\,{\bM}_{\{i_1,\ldots,i_k\}} 
	& = \left(\prod_{\ell=1}^{k}\delta_{i_\ell}\right)\left(1 -\frac{c}{2}\sum_{\ell=1}^k \frac{\alpha_{i_\ell}^2}{\delta_{i_\ell}}\right)~,\\
	& = \left(\prod_{\ell=1}^{k}\delta_{i_\ell}\right)\left(1 - \sum_{\ell=1}^k \frac{c_{{\rm sug},i_\ell}}{c}\right)~,\\
	& \geqslant 0~.
\end{split}
\end{equation}
%%%%%%
Whether this inequality represents a lower or an upper bound on $c$ depends on the details of the signs of the $\delta_i$, which in turn depend on whether the various flavor group factors contributing to the minor are sub-critical or super-critical. In particular, let us denote by $n^{i_1,\ldots,i_k}_{\rm sub}$ the number of subcritical factors in the minor. The inequality \eqref{eq:determinant_bound} then takes the form
%%%%%%
\begin{equation}\label{eq:general_bounds}
\begin{split}
\sum_{i\in I} c_{{\rm sug},i} \leqslant c &\qquad n^{i_1,\ldots,i_k}_{\rm sub}~{\rm odd}~,\\
\sum_{i\in I} c_{{\rm sug},i} \geqslant c &\qquad n^{i_1,\ldots,i_k}_{\rm sub}~{\rm even}~.
\end{split}
\end{equation}
%%%%%%
This parity dependence is quite powerful. We observe that if there are at least two subcritical factors in the flavor group, say $\gf_1$ and $\gf_2$, then by applying the above inequalities to the submatrices $\bM_{\{1,2\}}$ and $\bM_{\{1\}}$ we obtain a contradiction
%%%%%%
\begin{equation}\label{eq:bound_contradiction}
c \leqslant c_{{\rm sug},1} + c_{{\rm sug},2} < c_{{\rm sug},1} \leqslant c~,
\end{equation}
%%%%%%
where we've used the negativity of Sugawara central charges for subcritical Kac-Moody levels. Thus, compatibility of unitarity with VOA structures implies the following general principle:
\begingroup
\addtolength\leftmargini{-0.1in}
\begin{quote}
\emph{There can be at most one sub-critical simple factor in the flavor group of any unitary four-dimensional SCFT.}
\end{quote}
\endgroup
\noindent If this condition is obeyed and there exists a sub-critical factor, the remaining constraints all follow from the strongest bound, which arises from the minor that includes all non-critical factors in the flavor group and gives
%%%%%%
\begin{equation}\label{eq:sugawara_inequality}
c \geqslant c_{\rm sug, tot}~.
\end{equation}
%%%%%%
Alternatively, in the absence of a sub-critical factor, there are no additional constraints beyond the negativity of $c$ and $k_i$.

%%%%%%%%%%%%%%%%%%%%%%%%%%%%%%%%%%%%%%%%%%%%%%%%%%%%%%%%%%%%%%%%%%%%%%%%%%%%%%%%%%%%%%%%%%%%%%%%%%%%%%%%%%%%%%%%%%%%%

\vspace{-6pt}
\subsection{Bound saturation}
\label{subsec:saturation}

If there is a sub-critical flavor factor and the central charge takes its Sugawara value, thus saturating the bound \eqref{eq:sugawara_inequality}, then the corresponding zero eigenvector of ${\cM}$ describes a relation in the Higgs chiral ring,
%%%%%%
\begin{equation}\label{eq:Sugawara_null_vector}
\sum_{i=1}^n\tfrac{1}{2(k_i+h^\vee_i)}\mu^2_i = 0~.
\end{equation}
%%%%%%
Expressing $\mu^2_i$ in terms of $T_{{\rm sug},i}$ and $T_{\cT}$, this Higgs chiral ring relationship implies the equality of the true stress tensor and the (total) Sugawara stress tensor of the theory,
%%%%%%
\begin{equation}\label{eq:reductive_sugawara_saturation}
\sum_{i=1}^n T_{{\rm sug},i} = T_{\cT}~.
\end{equation}
%%%%%%
As such we find a natural generalization of the single-channel result,
%%%%%%
\begingroup
\addtolength\leftmargini{-0.1in}
\begin{quote}
\emph{If $c=c_{\rm sug, tot}$ in the associated VOA of a four-dimensional SCFT, then the true stress tensor is identified with the total Sugawara stress tensor}.
\end{quote}
\endgroup
%%%%%%
\noindent This must hold true as long as the stress tensor multiplet of the four-dimensional SCFT is the unique $\widehat{\cC}_{0(0,0)}$ multiplet in the theory.

%%%%%%%%%%%%%%%%%%%%%%%%%%%%%%%%%%%%%%%%%%%%%%%%%%%%%%%%%%%%%%%%%%%%%%%%%%%%%%%%%%%%%%%%%%%%%%%%%%%%%%%%%%%%%%%%%%%%%

\vspace{-6pt}
\subsection{Critical factors}
\label{sec:critical_factors}

Now we return to the general case, in which some simple factors in the global symmetry group are allowed to be at their critical level. The corresponding entries in $\Delta$ vanish, which allows us to derive two nontrivial results.

First, let us assume that there are at least two critical factors, call them $i$ and $j$. Then the $2\times2$ principal minor $\det\bM_{\{i,j\}}$ is manifestly zero since $\vec{\alpha}\otimes\vec{\alpha}$ has rank one. The corresponding zero eigenvector takes the simple form
%%%%%%
\begin{equation}\label{eq:critical_higgs_relations}
\cN_{ij}=\alpha_j\mu^2_i -\alpha_i\mu^2_j~.
\end{equation}
%%%%%%
Thus we find that the quadratic flavor singlets built from critical-level moment maps are all set equal to one another in the Higgs chiral ring. This phenomenon is familiar from the Higgs chiral rings of SCFTs of class $\cS$ with maximal punctures \cite{Benini:2009mz,Maruyoshi:2013hja}. However, here we see it as a simple consequence of unitarity and criticality of AKM levels, whereas previous derivations were somewhat more involved. The precise linear combination appearing in \eqref{eq:critical_higgs_relations} means that there is also a simple relation at the level of the associated VOA of Segal-Sugawara operators, namely,
%%%%%%
\begin{equation}\label{eq:Segal_Sugawara_null}
\alpha_jS_i = \alpha_iS_j~,
\end{equation}
%%%%%%
which has also featured prominently in the study of VOAs associated to class $\cS$ theories \cite{Beem:2014rza,Lemos:2014lua,Arakawa:2018egx}.

Next, we consider the case of a principal minor including one critical flavor factor (say $i$) and any number $k>0$ of non-critical factors (say $j_1,\ldots,j_k$). We find
%%%%%%
\begin{equation}\label{eq:critical_factor_minor}
\det\bM_{\{i,j_1,\ldots,j_k\}} = -\frac{c\,\alpha_i^2}{2}\prod_{n=1}^k\delta_{j_n}~.
\end{equation}
%%%%%%
Since $c$ is necessarily negative, unitarity requires that the product of the $\delta_{j}$ be positive. This must hold for any choice of the non-critical factors in the minor, so we learn a third general lesson:
%%%%%%
\begingroup
\addtolength\leftmargini{-0.1in}
\begin{quote}
\emph{In the presence of at least one critical flavor factor, there can be no sub-critical flavor factors}.
\end{quote}
\endgroup
%%%%%%
\noindent This completes the analysis of unitarity constraints arising from the positive semi-definiteness of $\bM$.

\vspace{-4pt}
\section{Summary and Discussion}
\label{sec:discussion}

Taken together, the results the we derived here lead to a surprisingly restrictive picture of the allowed structure of continuous flavor symmetries in unitary $\cN=2$ SCFTs with unique stress tensor multiplets. Here we summarize the final picture.
%%%%%%
\begin{itemize}
\item[(i)] If any critical simple flavor factors $\gf_i$ are present in the flavor group, then the corresponding ``quadratic Higgs singlet operators'' $\mu^2_i$ are all set equal by Higgs chiral ring relations.
\item[(ii)] In the presence of critical factors, there can be no sub-critical factors.
\item[(iii)] In the absence of critical factors, there can be at most a single sub-critical factor in the flavor group with $k_{i}+h^\vee_i > 0$. 
\item[(iv)] If there is a sub-critical factor, then the Virasoro central charge must obey $c \geqslant c_{\rm sug,tot}$.
\item[(v)] When this inequality is saturated, the true stress tensor in $T_\cT$ is given by the Sugawara stress tensor for the full flavor group of the theory.
%%%%%%
\end{itemize}
%%%%%%
Additionally, these results must hold when restricting to affine Kac-Moody subalgebras of the full flavor algebra. For example, in the rank-one $\ef_6$ Minahan-Nemeschansky SCFT, the critical-level $\suf(3)^3$ subalgebra inherits the Higgs chiral ring relations mentioned in $(iii)$ above.

\begin{table}[t]
\caption{\label{tab:KM_table}Unitarity bounds for Kac-Moody levels}
\begin{ruledtabular}
\begin{tabular}{@{\hspace{4em}} l l @{\hspace{4em}}}
\textrm{Flavor group factor}&
\textrm{Level bound}\\
\colrule
$\suf(2)$ & $k\leqslant -\tfrac13$\\
$\suf(n)$ & $k\leqslant -\tfrac{n}{2}$\\
\makebox[.55in][l]{$\sof(4)~,$}$n=4,\ldots,8$ & $k\leqslant -2$\\
\makebox[.55in][l]{$\sof(n)~,$}$n \geqslant 8$ & $k\leqslant 2-\tfrac{n}{2}$\\
\makebox[.55in][l]{$\uspf(n)~,$}$n\geqslant3$ & $k\leqslant-1-\tfrac{n}{2}$\\
$\gf_2$ & $k\leqslant-\tfrac53$\\
$\ff_4$ & $k\leqslant-\tfrac52$\\
$\ef_6$ & $k\leqslant-3$\\
$\ef_7$ & $k\leqslant-4$\\
$\ef_8$ & $k\leqslant-6$\\
\end{tabular}
\end{ruledtabular}
\end{table}

It is of some interest to examine these results in the context of the vast landscape of $\cN=2$ SCFTs that have been constructed to date, for example, as a part of class $\cS$. For instance, in \cite{Distler:2017xba,Distler:2018gbc} three-punctured fixtures in class $\cS$ were considered in light of whether or not they decompose as product SCFTs. One may immediately check that all of the examples in those papers can also be identified as product theories (\ie, as not having a unique stress tensor multiplet) by observing that they violate the single sub-critical factor rule. This is a slightly simpler criterion than that identified in the aforementioned references, which required the additional knowledge of the $c$ central charge for the theories in question.

The results of this letter should be complemented by the bounds obtained in \cite{Lemos:2015orc} (and generalized in \cite{Beem:2017ooy}) based on studying $\widehat{\cC}_{1(1/2,1/2)}$ multiplets appearing at level four in the Virasoro and affine Kac-Moody subalgebras of an associated VOA. These give rise to an upper bound for the Virasoro central charge
%%%%%%
\begin{equation}\label{eq:vir_c_upper_bound}
c \leqslant -\frac{11}{5}\left(1+\left(1+\frac{360}{121}\sum_{i=1}^n\frac{k_i d_i}{6k_i+h^\vee_i}\right)^{1/2}\,\right)~.
\end{equation}
%%%%%%
Thus, in the presence of a fixed sub-critical flavor factor, a finite interval of possible $c$-values exists, and the allowed interval shrinks upon the inclusion of additional factors in the flavor group. In addition, there are $\gf_i$-dependent upper bounds on the affine Kac-Moody levels $k_i$ which were derived in \cite{Beem:2013sza}, which we reproduce in Table \ref{tab:KM_table} for the reader's convenience. It may be of some interest to more thoroughly explore the space of allowed symmetries compatible with the full set of constraints we have uncovered here in conjunction with the previously established bounds given in \eqref{eq:vir_c_upper_bound} and Table~\ref{tab:KM_table}, similar to what was done in \cite{Lemos:2015orc}. We leave a thorough investigation of this type for the future.

\vspace{20pt}
\begin{acknowledgments}
I am very grateful to Jacques Distler, Madalena Lemos, Mario Martone, Wolfger Peelaers, and especially to Leonardo Rastelli for many helpful conversations on topics related to this work. I specifically thank Mario Martone and Jacques Distler for bringing the issue of Sugawara-type relations for reductive flavor groups to my attention in the first place, and to Mario Martone for comments on this manuscript. This work was supported in part by grant \#494786 from the Simons Foundation.
\end{acknowledgments}
\bibliography{flavor_bounds}
\end{document}